\documentclass[prb,preprint,showpacs,preprintnumbers,amsmath,amssymb]{revtex4}

\usepackage{graphicx}
\usepackage{dcolumn}
\usepackage{bm}
\usepackage{color}

\usepackage[mathscr]{eucal}

\preprint{BH formation: ver. 2.20}

\begin{document}

\title{Hydrogenation, dehydrogenation of $\alpha$-tetragonal boron and its transition to $\delta$-orthorhombic boron}

\author{Naoki Uemura and Koun Shirai$^{\dagger}$}
\email{koun@sanken.osaka-u.ac.jp}
\affiliation{Graduate School of Engineering, Nagoya Institute of Technology, Gokiso-cho, Showa-ku, Nagoya 466-8555, Japan \\
$^{\dagger}$Nanoscience and Nanotechnology Center, ISIR, Osaka University, 8-1 Mihogaoka, Ibaraki, Osaka 567-0047, Japan
}%

\author{Jens Kunstmann}
\affiliation{Theoretical Chemistry, Faculty of Chemistry and Food Chemistry \& Center for Advancing Electronics, TU Dresden, 01062 Dresden, Germany}

\author{Evgeny A. Ekimov, and Yuliya  B. Lebed$^{\ddagger}$}
\affiliation{Institute for High Pressure Physics, Russian Academy of Sciences, 142190, Troitsk, Russia \\
$^{\ddagger}$Institute for Nuclear Research, Russian Academy of Sciences, 142190, Troitsk, Russia}

\date{\today}

\begin{abstract}
Boron bulk crystals are marked by exceptional structural complexity and unusual related physical phenomena. Recent reports of hydrogenated $\alpha$-tetragonal and a new $\delta$-orthorhombic boron B$_{52}$ phase have raised many fundamental questions.
Using density functional theory calculations it is shown that hydrogenated $\alpha$-tetragonal boron  has at least two stable stoichiometric compositions, B$_{51}$H$_{7}$ and B$_{51}$H$_{3}$.
Thermodynamic modeling was used to qualitatively reproduce the two-step phase transition reported by Ekimov {\it et al.}~[J.~Mater.~Res.~\textbf{31}, 2773 (2016)]  upon annealing, which corresponds to successive transitions from B$_{51}$H$_{7}$ to B$_{51}$H$_{3}$ to pure B$_{52}$. 
The so obtained $\delta$-orthorhombic boron is an ordered, low-temperature phase and $\alpha$-tetragonal boron is a disordered, high-temperature phase of B$_{52}$. The two phases are connected by an order-disorder transition, that is associated with the migration of interstitial boron atoms. Atom migration is usually suppressed in strongly bound, covalent crystals. It is shown that the migration of boron atoms is likely to be assisted by the migration of hydrogen atoms upon annealing.
These results are in excellent agreement with the above mentioned experiment and they represent an important step forward for the understanding of boron and hydrogenated boron crystals. They further open a new avenue to control or remove the intrinsic defects of covalently bound crystals by utilizing volatile, foreign atoms. 
\end{abstract}


\maketitle

\section{Introduction}
\label{sec:intro}

Crystalline boron is the last monatomic system for which the phase diagram is not fully determined yet and even the number of phases is not certain. Currently only three, $\alpha$-rhombohedral (R), $\beta$-R, and $\gamma$-orthorhombic (O) boron, are generally accepted, but more than ten phases were reported before. However, the field underwent great progress, recently (for a review see Ref.~\onlinecite{Shirai17}). 

Among the various reported allotropes is $\alpha$-tetragonal boron ($\alpha$-T boron), which is a quite complex material. The $\alpha$-T phase was first synthesized in 1943.\cite{Laubengayer43} Its structure was thought to be B$_{50}$, where each unit cell is composed of four icosahedra and two interstitial atoms.\cite{Hoard51} But theorists questioned the existence of B$_{50}$ because the structure is short of 10 electrons to completely fill up its valence states.\cite{LH55}
Later, experiments showed that the apparent $\alpha$-T B$_{50}$ crystals were actually containing carbon and nitrogen impurities \cite{Amberger71, Ploog72, Will76} and the stability of these impurity-containing B$_{50}$C$_{2}$ and B$_{50}$N$_{2}$ crystals was also demonstrated by DFT calculations.\cite{Morrison92, Lee92a} 
However, Hayami and Otani theoretically showed that $\alpha$-T boron can be stabilized if the boron content is changed to B$_{52}$.\cite{Hayami10} At the same time several experimental works, mostly employing high-pressure high-temperature (HPHT) methods, reported the synthesis of $\alpha$-T B$_{52}$.\cite{Ekimov11,Ekimov11a,Qin12,Kurakevych12,Kurakevych13,Solozhenko13,Parakhonskiy13} 
The HPHT synthesis was not used in the early days, and it is therefore likely that $\alpha$-T boron is thermodynamically stable at high pressure and high temperature.\cite{Oganov09} 
Unfortunately, owing to the lack of accurate information about the crystal structures, it is not clear if the various reported samples represent the same phase or if they contain impurities. 
A recent theoretical study by Uemura {\it et al.} has identified the characteristic features of pure $\alpha$-T boron B$_{52}$.\cite{Uemura16} It was shown that the occupation of interstitial sites as well as non-stoichiometric compositions, {\it i.e.}~non-integer number of atoms per primitive unit cell, are crucial for the stability of the system. Their results could then be used to define a family of $\alpha$-T boron systems. \cite{Ekimov11}

Here we are concerned with hydrogenated $\alpha$-T boron and its relation to a new orthorhombic boron phase. Hydrogenated $\alpha$-T boron crystals, with compositions B$_{51.5}$H$_{m}$, were prepared by Ekimov {\it el al}.\cite{Ekimov11,Ekimov16} Their samples were synthesized by thermal decomposition of decaborane at high temperature ($T \sim 1100-1300^{\circ}$C) and high pressure ($p = 8-9$ GPa) and the initial H content was $m=7.7$. This is exceptionally high for a crystalline semiconductor.  Afterwards they annealed the crystals at ambient pressure and observed two phase transitions as $T$ was raised. At $T=450^{\circ}$C, $m$ reduced to $m=4.7$ and at $T=700^{\circ}$C, hydrogen was fully released. Interestingly, during the second step the tetragonal lattice underwent a transition to an orthorhombic lattice. 
In the following we will call this structure $\delta$-O boron in order to distinguish it from the more established $\gamma$-O boron. The formal composition B$_{n}$ of $\delta$-O boron is $n=52$, although X-ray analysis showed $n$ to be in the range from 51.6 to 52. This structure was theoretically predicted by Hayami and Otani \cite{Hayami10}, and Zhu.\cite{Zhu12} However, their calculations were limited to the primitive unit cell and therefore miss two important aspects of  boron crystals which were found to be crucial for the stability of $\alpha$-T boron: the effect of disorder and non-stoichiometric compositions.\cite{Uemura16} Similar investigations on $\delta$-O are necessary before it can be considered as a new boron allotrope.

In this paper, the structure of hydrogenated $\alpha$-T boron, the process of dehydrogenation, and the phase transition to $\delta$-O boron are studied theoretically. The study of hydrogenation/dehydrogenation is of general importance for material research, and our results are potentially useful for hydrogen-related technologies, such as hydrogen storage.\cite{Berg08,Nishino17} Our insights on the phase transition are of central importance for ongoing efforts to create the phase diagram of boron.
The paper is organized as follows. Section \ref{sec:method} describes the crystal structure and calculation methods, along with several definitions used throughout this paper.
In Sec.~\ref{sec:structure}, a structural study on hydrogenated $\alpha$-T boron is given.
In Sec.~\ref{sec:anneal}, the process of dehydrogenation is studied. In Sec.~\ref{sec:phase-transition}, the phase transition and its implications are discussed. Finally a summary is given in Sec.~\ref{sec:conclusion}.

\section{Crystal structure, definitions, and method}
\label{sec:method-structure}
\subsection{Crystal structure}
\label{sec:struct}
\begin{figure}[htpb]
\centering
\includegraphics[width=7.0 cm]{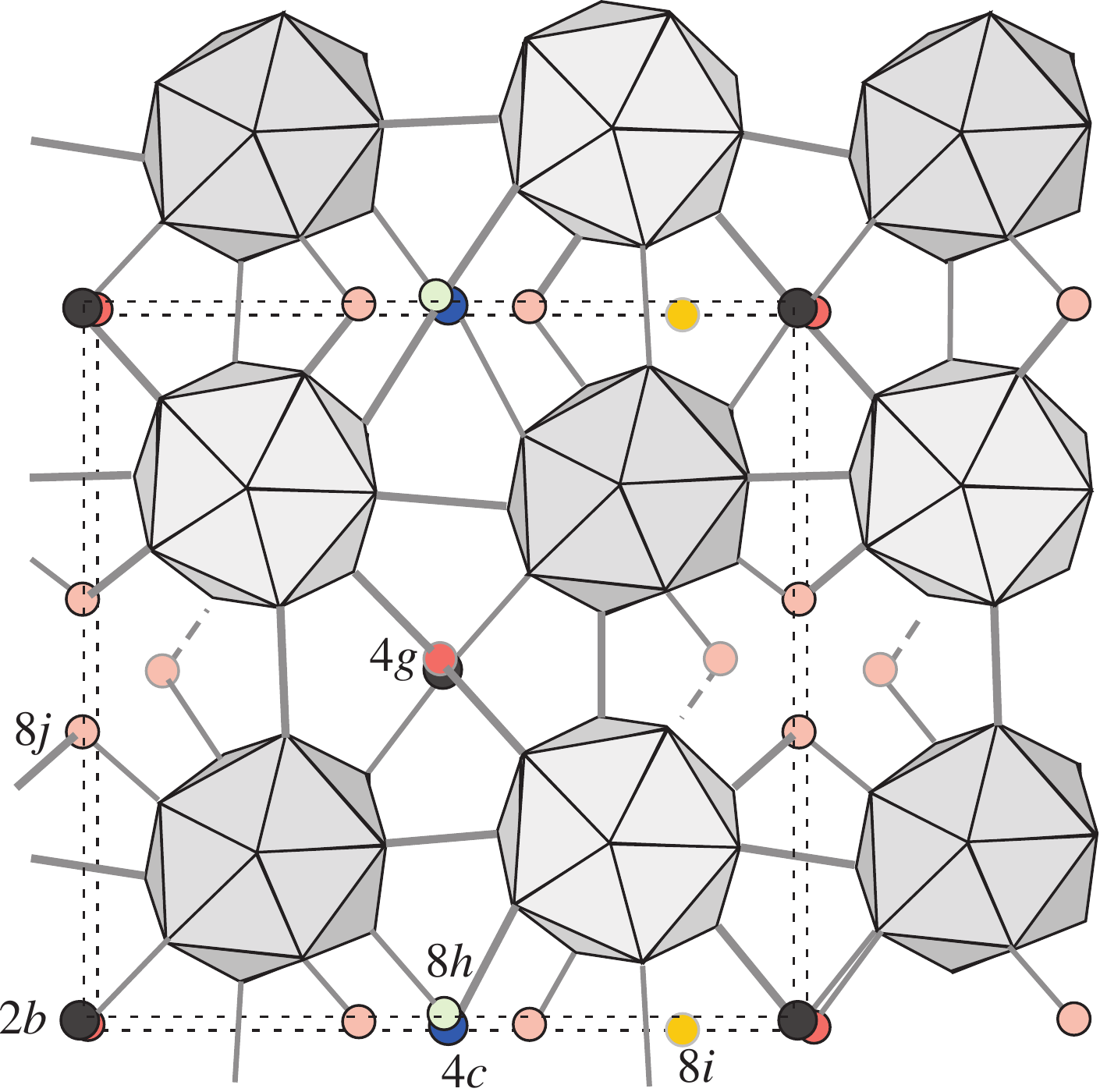}
\hspace{1.0 cm}
\includegraphics[width=8.0 cm]{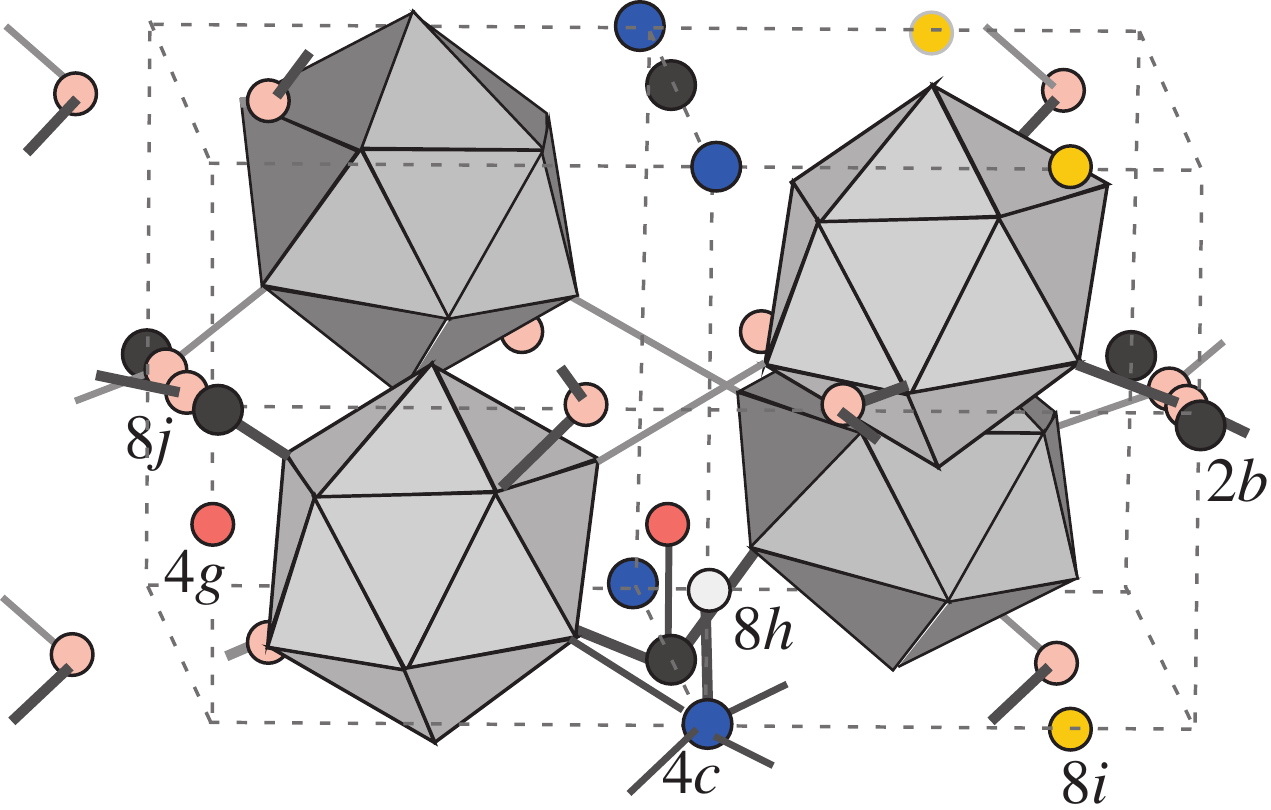}
\caption{The atomic structure of $\alpha$-tetragonal boron crystals and the position of interstitial sites.
(left)/(right) Top/side views of the unit cell. The interstitial sites for boron atoms are (2\textit{b}) (black) and (4\textit{c}) (blue). The interstitial sites for hydrogen atoms are (4\textit{g}) (red), (8\textit{h}) (white), (8\textit{i}) (yellow), and (8\textit{j}) (pink). The names of Wyckoff positions are according to the high-symmetry space group P4$_{2}$/nnm: $(2a) (000)$, $(2b) (00 \frac{1}{2})$, $(4c) (0 \frac{1}{2}0)$, $(4g) (00z)$, $(8h) (0\frac{1}{2}z)$, $(8i) (x 0 0)$,
$(8j) (x 0 \frac{1}{2})$. }
\label{fig:xtl}
\end{figure}

The basic unit cell of $\alpha$-T-type crystals contains four icosahedra and two interstitial atoms (B$_{50}$). 
To ease comparison with the idealized $\alpha$-T B$_{50}$ in previous works, we describe our systems within the high-symmetry space group $P4_{2}/nnm$, although the real structures have lower symmetry. 
There are several interstitial sites in $\alpha$-T boron and they are illustrated in Fig.~\ref{fig:xtl}. The site names are given by their Wyckoff notations in parenthesis, such as $(2b)$ or $(4c)$. Among them, the $(2b)$ site is a fully occupied site (FOS) and the occupying atoms can be considered as a part of the B$_{50}$ host crystal. Others are partially occupied sites (POS), which could also be considered as defects.


\begin{figure}[htpb]
\centering
\includegraphics[width=140 mm,bb=0 0 841 368]{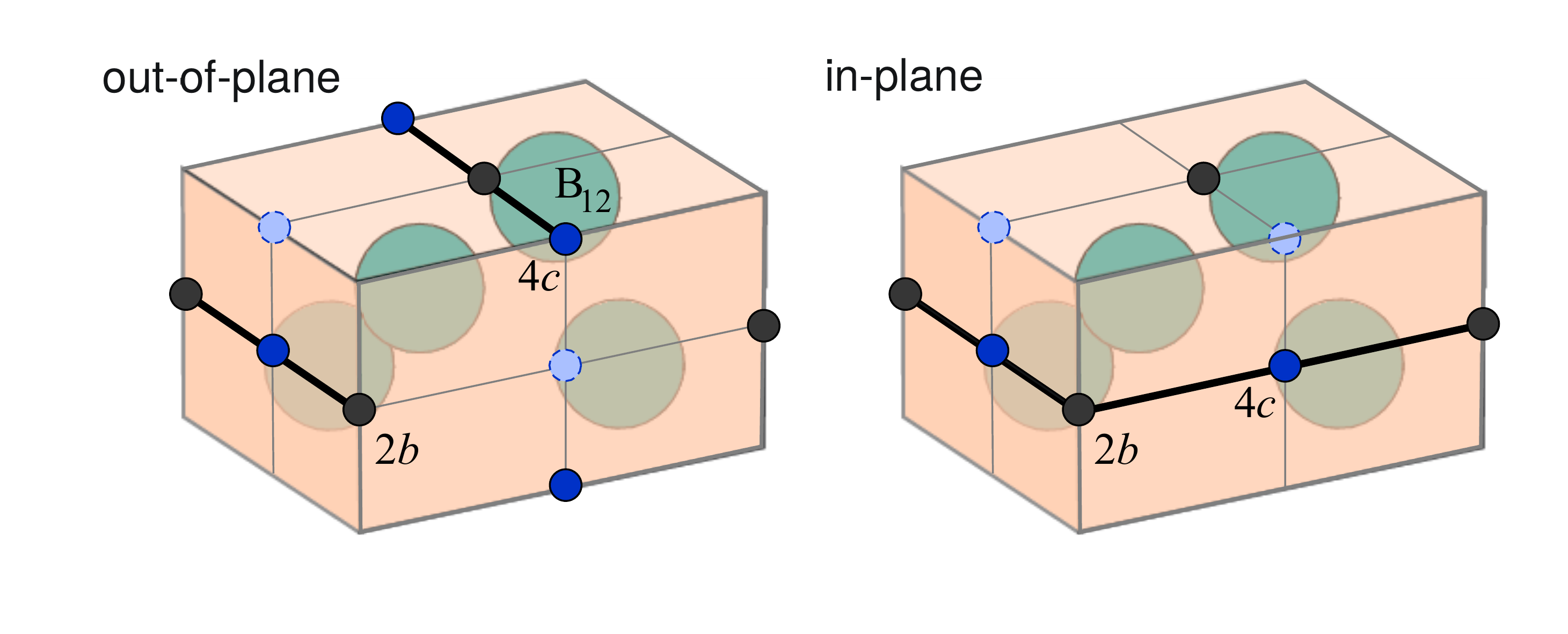}
\caption{Out-of-plane and in-plane arrangements of two $(4c)$-site atoms in $\alpha$-T boron B$_{52}$ and their relation to the orthorhombic distortion of the unit cell.
Occupied $(4c)$-sites are dark blue, unoccupied ones are indicated by light blue. The two $2b$ sites (black) are fully occupied. Large spheres inside unit cells represent B$_{12}$ units.
Thick lines represent the formation of covalent bonds along (left) one crystallographic direction for the out-of-plane configuration, leading to a lattice with orthorhombic symmetry and (right) two directions for the in-plane configuration, leading to tetragonal symmetry.
The locations of the $(4c)$ sites are $(0, 1/2, 0), \ (0, 1/2, 1/2), \ (1/2, 0, 1/2), \ (1/2, 0, 0)$. }
\label{fig:zig-zag}
\end{figure}

The lowest-energy structure of 
pure $\alpha$-T boron is B$_{52}$, i.e., the basic B$_{50}$ structure with two additional $(4c)$-site atoms.
\cite{Hayami10,Uemura16}
There are four symmetry-equivalent sites for $(4c)$ in $\alpha$-T boron. When two atoms occupy two $(4c)$ sites in the same $ab$ plane, we call it the {\it in-plane} configuration.
When they occupy two $(4c)$ sites in different $ab$ planes, we call it the {\it out-of-plane} configuration (see Fig.~\ref{fig:zig-zag}).
The lowest-energy structure of pure $\alpha$-T boron is B$_{52}$ in the out-of-plane configuration\cite{Hayami10,Uemura16}
and in this paper pure $\alpha$-T boron is always related to this structure, unless otherwise stated.

The formula unit of hydrogenated $\alpha$-T boron is B$_{n}$H$_{m}$; henceforth, $n$ and $m$ are used to indicate B and H contents, respectively. 
Unfortunately, the chemical composition, given by $n$ and $m$, is not quite certain.
When Ekimov {\it et al}.~reported the synthesis of pure $\alpha$-T boron, $n=51.5$ was specified.\cite{Ekimov11} However, hydrogen inclusion was later discovered by mass spectroscopy, and  two kinds of compositions BH$_{0.15}$ and BH$_{0.09}$ were found by annealing. Furthermore, the B content varied from 51.5 to 52.0 during the annealing.\cite{Ekimov16} Whether or not the difference in the B content is due to experimental error is unclear. If we assume $n=51.5$ for the parent crystal, then BH$_{0.15}$ and BH$_{0.09}$ correspond to $m=7.7$ and $4.6$, respectively. These are the compositions mentioned in the introduction. 

\subsection{Computational methods}
\label{sec:method}
The electronic structures of the considered systems were studied by density functional theory (DFT) using the pseudopotential method and the Osaka2k code.\cite{Osaka2k} It uses the parameterization by Perdew and Zunger for the local density approximation (LDA)\cite{PZ81}, the Perdew-Burke-Ernzerhof form of the generalized gradient approximation (GGA) \cite{PBE96} and Troullier-Martins pseudopotentials \cite{TM91} with the fully separable Kleinman-Bylander form \cite{KB82}.
Various $k$-point sampling methods were used. For calculations using the primitive unit cell, a $2 \times 2 \times 4$ Monkhorst-Pack mesh
was used, and for supercell calculations the $\Gamma$-point only. 
In all the cases, the kinetic cutoff energy $E_{\rm cutoff}$ was 40 Ry.
The convergence was well tested in our previous studies \cite{Uemura16} and increasing $E_{\rm cutoff}$ to 80 Ry changes the formation energy only by 2 to 4 meV/atom.
Structural optimizations were performed with respect to atomic positions and cell parameters and no constraints on the crystal symmetry were imposed, except the tetragonal symmetry for the lattice parameters. 
GGA was used only for the determination of the lattice parameters. 
Although GGA usually improves formation energies, it is not the case for systems with hydrogen.\cite{Perdew03} 

\subsection{Thermodynamic models}

Experimentally, hydrogenated $\alpha$-T boron was prepared by decomposition of decaborane (B$_{10}$H$_{14}$) at high pressure. It is difficult to simulate the reaction exactly, because the decomposition process of borane is very complicated. For basic thermodynamic considerations, it is, however, useful to consider the formation reaction
\begin{equation}
\frac{n}{52} {\rm B}_{52} + m \ {\rm H}^{(s)} \rightarrow {\rm B_{n}H}_{m}, 
\label{eq:react-hipres}
\end{equation}

where B$_{52}$ is assumed to be in the lowest-energy structure of pure $\alpha$-T $\mathrm{B}_{52}$\cite{Uemura16} and solid hydrogen ${\rm H}^{(s)}$ of $Pa3$ crystal symmetry \cite{Mao94,Cui95} is considered as the starting material to emulate high-pressure conditions.
Thus, the total formation energy $E_{f}$ of B$_{n}$H$_{m}$ is 
\begin{equation}
  E_{f} [\mathrm{B}_{n} \mathrm{H}_m] = E [\mathrm{B}_{n} \mathrm{H}_m] - \biggl\{  \frac{n}{52}  E[\mathrm{B}_{52}]  + m E[\mathrm{H}^{(s)}] \biggr\}
\label{eq:formationE}
\end{equation}

and $E[A]$ is the DFT total energy of a system $A$.
By considering reaction (\ref{eq:react-hipres}), the H content may be overestimated because the chemical potential of H is a maximum for solids. On the other hand, the B content may be underestimated, since the intermediate products of decomposition of B$_{10}$H$_{14}$ are less stable than the B$_{52}$ solid.

In the discussion of hydrogen inclusion below, the individual hydrogen atoms can be considered as impurity atoms. Then the total formation energy $E_{f}$ indicates the energy cost to add the impurity atom to the host crystal $\mathrm{B}_{52}$. Alternatively one can consider the B$_{n}$H$_{m}$ crystals as compounds with varying compositions. To enable the comparison between them, it is common to consider the normalized formation energy 
\begin{equation}
  e_f = \frac{E_{f}}{n+m}.
\label{eq:formEnorm}
\end{equation}

Now, let $j$ refer to a specific configuration of H atoms. 
The experimental observed hydrogen content is the ensemble (thermal) average $\langle m \rangle$, where different configurations $j$ contribute to $\langle m \rangle$ through their formation energies $E_{f,mj}$.
The ensemble average is obtained through calculating the partition function ${\cal Z}$,
\begin{equation}
{\cal Z} = \sum_{m} \sum_{j}{g_{mj} \exp(-\beta E_{f,mj}) },
\label{eq:gz}
\end{equation}
where $g_{mj}$ is the multiplicity of the $j$-th configuration of the H content $m$ and $\beta=1/k_\mathrm{B}T$ is inverse to temperature with the Boltzmann constant $k_\mathrm{B}$.
In Eq.~(\ref{eq:gz}), the configurations $j$ for a given value of $m$ are grouped and the summation over a group is denoted by $z_{m}$,
\begin{equation}
{\cal Z} = \sum_{m}{z_{m}}
 = \sum_{m} {\exp( -\beta F_{m})},
\label{eq:szn}
\end{equation}
where $F_{m}$ is the partial free energy for the H content $m$. The entropic contributions to $F_{m}$ are then entirely from the configurational degrees of freedom. 
Finally, the mean value $\langle m \rangle$ is obtained by
\begin{equation}
\langle m \rangle = \frac{1}{{\cal Z}} \sum_{m} m \ {\exp( -\beta F_{m})}.
\label{eq:mvalue}
\end{equation}

In order to study the annealing of B$_{n}$H$_{m}$ at ambient pressure, Eq.~(\ref{eq:react-hipres}) is modified such that solid hydrogen H$^{(s)}$ is replaced by molecular hydrogen in the gas phase $1/2$ H$_2^{(g)}$ and its partial pressure $p$ is a parameter.
As the density of hydrogen gas is much smaller than that of the solid, we have to replace the total energy $E[\mathrm{H}^{(s)}]$ in Eq.~(\ref{eq:formationE}) by the chemical potential of hydrogen gas $\mu_{\rm H_{2}}$.
For $\mu_{\rm H_{2}}$, we use the ideal gases expression
\begin{equation}
\mu_{\rm H_{2}} = R T \left[ (c+1) - \ln( T^{c} v ) \right],
\label{eq:muofH}
\end{equation}
where $c=5/2$, $R$ is the gas constant, and $v$ is the molecular volume. Real gas corrections to Eq.~(\ref{eq:muofH}) (van der Waals or the inclusion of the latent heat of phase transitions) are estimated to be negligible for the considered energy scales.
The $mj$-th component of the free energy $F_{mj}(T,p)$ of the $j$-th H configuration and $m$ H atoms is then obtained by 
\begin{equation}
F_{mj}(T,p) = E_{f, mj} - \frac{m}{2} \mu_{\rm H_{2}}(T,p)
\label{eq:formF}
\end{equation}
For the determination of the partition function ${\cal Z}$ in Eq.~({\ref{eq:gz}}), this free energy $F_{mj}(T,p)$ is used instead of $E_{f,mj}$ and $f=F/(n+m)$ indicates the normalized free energy.

\section{Properties of hydrogenated $\alpha$-T boron}
\label{sec:structure}

\subsection{Bonding and vibrational properties of interstitial hydrogen}
\label{sec:bonding-POS}

For pure $\alpha$-T boron the characteristics of interstitial sites were analyzed previously.\cite{Uemura16}
It was found that up to a composition $n=52$ the $(4c)$ site is the most preferable POS, followed by $(8h)$ and $(8i)$. The occupation of $(4g)$ is negligible and $(2a)$ is the least preferable one (however, this site is important for the inclusion of N or C atoms).

The site occupancies of hydrogenated $\alpha$-T boron were measured by Ekimov {\it et al}.\cite{Ekimov11} Since H atoms cannot be detected by X-ray diffraction, only the occupancies of B atoms were measured to be: 100\% ($2b$), 31\% ($4c$), 6\% ($4g$). An apparent feature of hydrogenated $\alpha$-T boron is the occupation of the $(4g)$ site, which has never been reported before for pure $\alpha$-T boron.

\begin{table}
 \caption{Total $E_{f}$ and normalized $e_{f}$ formation energies of B$_{n}$H for various H sites, according to Eqs.~(\ref{eq:formationE}) and (\ref{eq:formEnorm}), respectively. A low energy indicates a preferable site. For B$_{51}$H, an extra B atom is located at a $(4c)$ site. The center of B$_{12}$ is denoted as (ico).}
 \label{table:Hf_B50H}
 \begin{ruledtabular}
  \begin{tabular}{c| cc| cc}
    \multicolumn{1}{c|}{H site}  & \multicolumn{2}{c|}{$E_{f}$ [eV]} & 
    \multicolumn{2}{c}{$e_{f}$ [meV/atom]} \\ \cline{2-5}
        & \multicolumn{1}{c}{B$_{50}$H} & \multicolumn{1}{c|}{B$_{51}$H} &
        \multicolumn{1}{c}{B$_{50}$H} & \multicolumn{1}{c}{B$_{51}$H} \\
         \hline
     (2\textit{a})                 & 4.42 & 2.47 & 88.4 & 48.4 \\
     (4\textit{c})                 & 2.53 & 0.45 & 50.6 & 8.8 \\
     (4\textit{g})                & 3.80 & 1.83 & 76.0 & 35.9 \\
     (8\textit{h})                 & 2.57 & 0.49 & 51.4 & 9.6 \\
     (8\textit{i})                 & 2.46 & 0.45 & 49.2 & 8.8 \\
     (8\textit{j})                  & 2.02 & 0.02 & 40.4 & 0.4 \\
     (ico) & 5.79   & 3.58  & 115.8 & 70.2 \\  
 \end{tabular}
\label{tab:inter-sites-B5n}
\end{ruledtabular}
\end{table}

In order to determine the preferable hydrogen sites, we calculated the formation energies as listed in Table \ref{tab:inter-sites-B5n}. 
For all H sites, $e_{f}$ decreases when the B content increases from 50 to 51. We will explain this results in the discussion below.
Let us first focus on the B$_{51}$H case, where the interstitial B atom is located at a $(4c)$ site.
$e_{f}$ of the $(8j)$ site is by far smallest, which is interesting because $(8j)$ is not very relevant for pure $\alpha$-T boron. The next most preferable sites are $(4c)$, $(8i)$, and $(8h)$. 
For the $(4g)$ site $e_{f}$ is quite high (35.9 meV/atom). However, the value is significantly reduced from that of pure $\alpha$-T boron (76 meV/atom) \cite{Uemura16} so that we cannot fully ignore this cite.
The site at the center of the icosahedron (ico) can be excluded from further consideration because  $e_f$ is far too high.

\begin{table}
\caption{Boron-hydrogen bond lengths in B$_{50}$H, reflecting the bonding environment of a crystallographic site. A favorable B-H bond length is ca.~1.2 \AA. Expressions like $2 \times 1.69$ mean two bonds with a length of 1.69 \AA.}
\begin{ruledtabular}
  \begin{tabular}{cccccccc}
    Site        & (2\textit{a})    & (4\textit{c})              & (4\textit{g}) & (8\textit{h}) & (8\textit{i}) & (8\textit{j}) & (Ico.) \\ \hline
    N. of bonds         & 8                 & 4                           & 3              & 4              & 4             & 2             & 12                          \\ \hline
    bond length [\AA]  & 2.08 - 2.22 &  2 ${\times}$ 1.69 & 1.30         & 1.63          & 1.64       &1.33         & 1.58 - 1.79             \\ 
                                         & (ave. 2.15) &  2 ${\times}$ 1.75 & 1.58         & 1.71          & 1.69       &                & (ave. 1.72)             \\
                                         &                   &                              & 1.79         & 1.76          & 1.73       &                &                               \\
                                         &                   &                              &                 & 1.83          & 1.78       &                &                               \\ 
 \end{tabular}
 \end{ruledtabular}
 \label{tab:bond-length}
\end{table}

These differences in the formation energy arise from the different bonding environments of interstitial sites in B$_{50}$H and B$_{51}$H. 
Table \ref{tab:bond-length} lists the lengths of H$-$B bonds up to third nearest neighbor atoms. 
As stated before, the $(8j)$ site is the most preferable one for H atom inclusion (bond length is 1.33 {\AA}). Its bonding environment is best described in analogy with the diborane molecule B$_{2}$H$_{6}$,\cite{Pauling} which has a B$-$H$-$B bridge structure with the bond length 1.33 {\AA} and the bond angle 83$^{\circ}$. As seen in Fig.~\ref{fig:xtl} (left), $(8j)$ H atoms form similar B$-$H$-$B bridges that link two adjacent icosahedra and the B$-$H bond length of 1.33 is identical to that of B$_{2}$H$_{6}$. Thus, $(8j)$ sites are the best ones for H incorporation. Similar B$-$H$-$B bridge structures were reported before for hydrogen in $\alpha$-R boron.\cite{Wagner11} 
The distance from an $(8i)$ site to an apex atom of nearest B$_{12}$ (1.64 \AA) is a typical B$-$B bond length. Therefore, the site is best occupied by B atoms. 
The $(4g)$ site is not very favorable for forming B$-$H bonds. However, as will become clearer below, it plays an important role in hydrogenated $\alpha$-T boron.
At a $(2b)$ site the distance to the neighboring B atom is mostly 1.30 {\AA} and  1.23 {\AA} in a few cases. A charge density analysis around this site (see Supplemental Material S.1) indicates it to be a single bond. 
B$_{2}$H$_{6}$ is terminated by B$-$H single bonds with bond lengths of 1.19 {\AA}. Our bond length of 1.30 {\AA} is much longer. As a consequence, the vibrational properties are different in the two cases. 
In the ${\rm B_{2}H_{6}}$ molecule, the vibration of the terminal B$-$H bond is an IR-active stretching mode at 2558 ${\rm cm^{-1}}$.\cite{Bell45,Brown55} However, our phonon calculations of $\alpha$-T B$_{51}$H$_{m}$ show that the B$-$H bonds at $(4g)$ sites usually have stretching modes at about 2000 ${\rm cm^{-1}}$. 
Only when the bond length is shorter than 1.23 \AA, the mode has a high frequency about 2500 ${\rm cm^{-1}}$. This explains why hydrogen was initially not detected by IR spectroscopy. 
Furthermore, the ${\rm B_{2}H_{6}}$ molecule has an IR-active B$-$H$-$B mode at 1608 ${\rm cm^{-1}}$. In $\alpha$-T B$_{51}$H$_{m}$ the frequency of this mode  ranges from 1600 to 2000 ${\rm cm^{-1}}$, depending on the H configurations. These phonon properties are thus useful for the identification of hydrogen and its configuration in $\alpha$-T B$_{51}$H$_{m}$. 



\subsection{Electronic structure - band filling}
\label{sec:band-filling}
\begin{figure}[htpb]
\centering
\includegraphics[keepaspectratio, scale=0.4]{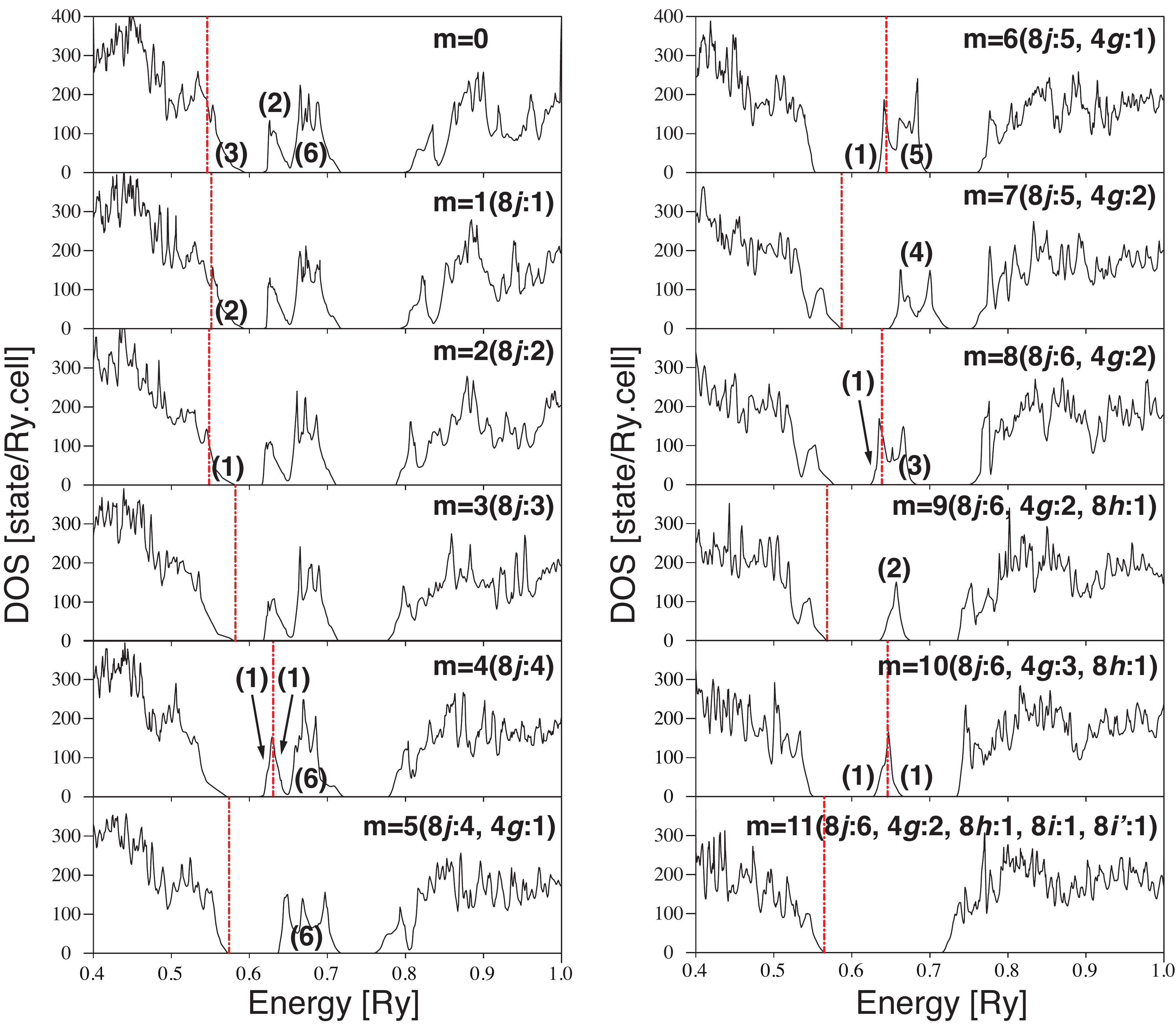}
\caption{Band filling through hydrogen as indicated by the density of states of B$_{51}$H$_{m}$. In each panel the number of spin states are indicated in parentheses. The number of hydrogen atoms $m$ and their location are given in the upper right corners of the panels; here $(8j:3)$ means three H atoms at $(8j)$ sites. The red, vertical line is the Fermi energy. Systems with completely filled valence bands are ($m=3,5,7,...$) are particularly stable.}
\label{fig:dos}
\end{figure}

Next, let us investigate how the electronic structure evolves when changing the number of H atoms $m$.
A series of density of states (DOS) plots for hydrogenated $\alpha$-T boron are shown in Fig.~\ref{fig:dos}, where the number of B atoms is $n=51$ (experimental value $n=51.5$). The H atoms are successively added, mostly at the $(8j)$ sites because $e_{f}$ is small. The DOS plots represent minimum-energy configurations for each $m$.

The B$_{51}$ system (top left panel of Fig.~\ref{fig:dos}) is short of three electrons to completely fill up the valence band and eight empty states (four bands) appear in the band gap.\cite{Uemura16} The character of two of these in-gap bands corresponds to $p$-orbitals of $(2b)$-site B atoms that are oriented along the $c$-axis. The remaining two bands have $p$-orbital character of the atom at the $(4c)$ site.
As $m$ increases up to $m=3$, the three unoccupied states of the valence bands are successively filled, with the gap states being unchanged. This is contrast to the evolution of the DOS in pure boron, where placing B atoms at interstitial sites creates new in-gap states.\cite{Uemura16} At $m=3$, the valence band is completely filled, which renders ${\rm B_{51}H_{3}}$ to be a stable structure. 
However, eight in-gap states still remain, which are filled one by one as the H content is further increased. 
Interestingly, an in-gap band is moved from the gap into the valence band, whenever two H atoms are added. 
This means that these two electrons either form a B$-$H$-$B bridge bond through a (8j)-site or a B$-$H bond on top of a (2b)-site.
In this way, the in-gap states are eliminated one by one until eventually all of them are removed for $m=11$. Thus the DOS analysis identified a series of potentially stable structures for $m=3,5,7,9,11$. Their stability is confirmed by low formation energies, as discussed in the next section and as discernible by the energy minima in Fig.~\ref{fig:B51H}.

Hydrogenation could thus be a practical method to control the in-gap states of $\alpha$-T boron, which cannot be achieved by varying the B content only, as explained above. Furthermore, it could be possible to identify these states by optical spectroscopy. Since the number of in-gap states in pure and hydrogenated $\alpha$-T boron differs, optical spectroscopy could help to distinguish corresponding crystals (another method is to measure the lattice parameters, as discussed in Sec.~\ref{sec:volume}).

\subsection{Stability of hydrogenated $\alpha$-tetragonal boron}
\label{sec:formation-energy}
In order to learn about the energetic stability and the thermodynamic properties of hydrogenated $\alpha$-T boron, we have examined various compositions of B$_{n}$H$_{m}$, for $n=50, \dots, 52$ and $m=0, \dots, 12$ (for the complete data (see Supplemental Material S.2).
For every composition, there are many atomic configurations $j$, however, many can be ignored because they have high formation energies and do not contribute to the partition function Eq.~(\ref{eq:gz}) at low temperatures. For example, sites that are near a place where a B atom forms a covalent bond can be omitted because the addition of a H atom weakens the bond. Furthermore, as seen in Table \ref{tab:inter-sites-B5n}, interstitial sites such as $(8j)$ are energetically preferred over others.

>From the insights about the valence band filling in the previous section and in similar discussions in Ref.~\onlinecite{Uemura16}, we expect that for B$_{52}$ the addition of H atoms is not energetically favorable because the valence band is fully occupied. In fact, the present calculations confirm this and the formation energies $e_f$ of B$_{52}$H$_{m}$ show a monotonic increase with increasing $m$ (see Supplemental Material S.2). 
On the other hand, for $n=50$ and $n=51$ $e_f$ is expected to decrease as $m$ increases. Furthermore, the rigid-band model predicts an energy minimum at $m=7$ for $n=51$ and at $m=10$ for $n=50$. Our calculations confirm both expectations (see Supplemental Material S.2).

Comparing all compositions, the absolute minimum of $e_f$ is found for $n=50$ and $m=10$. However, the minimum of the $n=51$ series at $m=7$ is much closer to the experiment, where $n=51.5$ and $m=7.7$ were reported. The reasons for this discrepancy could be our oversimplified assumptions for the reaction (\ref{eq:react-hipres}) that may underestimate $n$. 
Therefore, in the following we do not seek the exact value of $n$ but rather focus on the variation of $m$, by assuming $n=51$.

\begin{figure}[htpb]
\centering
\includegraphics[width=0.8\columnwidth]{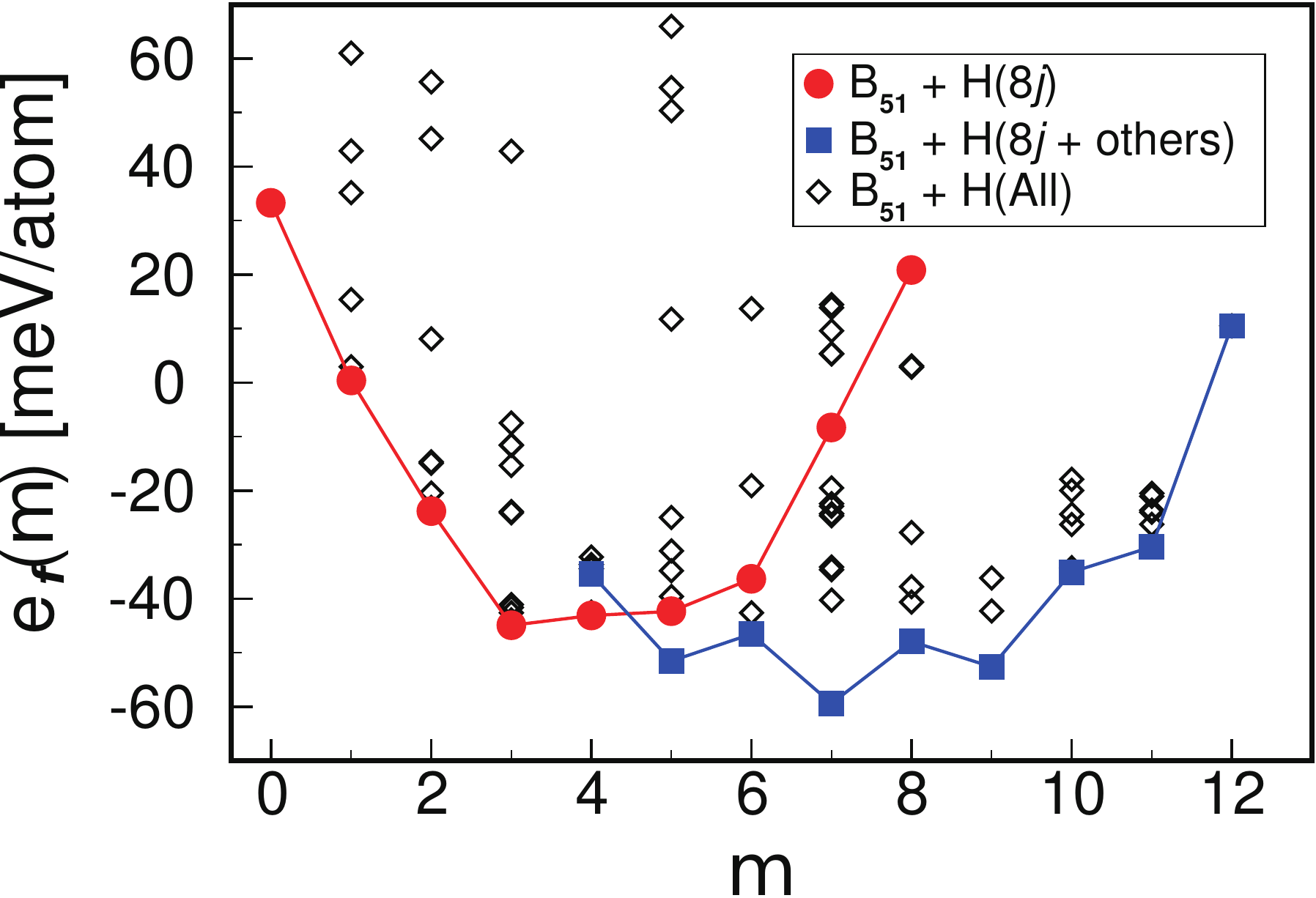}
\caption{Formation energies $e_f$ of B$_{51}$H$_{n}$ for varying H content $m$. Low energies  indicate stable structures; the most stables ones are found for $m=3,5,7,\dots$
The lowest-energy configurations for each $m$ are connected by red and blue lines. Red circles indicate hydrogen configurations that only occupy $(8j)$ sites,  blue squares indicate configurations that include more sites. }
      \label{fig:B51H}
\end{figure}

The formation energies for B$_{51}$H$_{m}$ are shown in Fig.~\ref{fig:B51H}.
Red and blue lines indicate the lowest-energy configurations for each $m$. 
The red line represents the configurations that only occupy $(8j)$ sites. Up to $m=4$, the lowest-energy configurations are found in this series. 
As expected from the analysis in Sec.~\ref{sec:bonding-POS}, the $(8j)$ sites are the most preferential ones for H occupation, and hence this result is reasonable. Within this range the lowest-energy configuration is $m=3$ (three $(8j)$-sites are occupied). As seen in Fig.~\ref{fig:dos}, this corresponds to completely filled valence bands, with leaving the in-gap states unoccupied.

For $m>4$, other sites come to participate, although $(8j)$-sites are still major sites. This series is indicated by a blue line. The lowest-energy state is found at $m=7$ and the H configuration is: five $(8j)$ and two $(4g)$ sites (also see Fig.~\ref{fig:dos}). 
Interestingly, the next preferential sites to occupy after $(8j)$ are $(4g)$ and not $(8h)$ or $(8i)$ sites, even though the low formation energies in Table \ref{tab:inter-sites-B5n} suggest that $(8h)$ and $(8i)$ sites should follow. 
The reason for this is found by using the empirical rule from above, namely, placing a H atom close to a site where covalent bonds are already formed causes $e_f$ to increase. When $(8j)$ sites are already occupied, additional $(8h)$ or $(8i)$-site occupation increases $e_f$ because the $(8h)$ and $(8i)$ sites are close to $(8j)$ site. Therefore $(4g)$-sites are more favorable to be occupied in that situation.

A notable observation in Fig.~\ref{fig:B51H} is that there are two local minima at $m=5$ and $m=7$ that nicely correspond to the experimentally observed H content of the crystals for which a phase transition occurred in the annealing experiment, namely 4.6 and 7.7.\cite{Ekimov16} This correspondence is discussed in more detailed in the next section. 

With these results we can calculate the thermodynamic average $\langle m \rangle$ at finite temperatures, as given by Eq.~(\ref{eq:mvalue}). Using the energies of all B$_{51}$H$_{m}$ configurations we obtain $\langle m \rangle =  7.1$ at $T=2000$ K, which is in good agreement with the experimental value $m=7.7$.
If all configurations of B$_{50}$H$_{m}$ are also taken into account $\langle m \rangle$ increases to $8.7$. 
However, in this case the B content $\langle n \rangle$ decreases and the agreement with respect to the experimental B content is less good. Therefore and for reasons given above, we fix the B content to $n=51$.

Figure \ref{fig:B51H} also allows to determine the chemical potential of hydrogen $\mu_{\rm H}^{(s)} = (\partial E_{f}/\partial m)_{m \rightarrow 0}$ in solid $\alpha$-B. From the slope of  $E_f(m)$ $\mu_{\rm H}^{(s)}$ is found to be $-1.5$ eV/(H atom). This value can be compared with $-3.5$ eV/(N atom) for B$_{50}$N$_{2}$ and $-0.5$ eV/(C atom) for B$_{50}$C$_{2}$ (these values were obtained from Ref.~\onlinecite{Uemura16}, though they are not explicitly given there). Hence, the affinity of hydrogen to $\alpha$-T boron is strong and is in intermediate between C and N. 

\subsection{Volume change}
\label{sec:volume}

\begin{figure}[htbp]
      \centering
      \includegraphics[keepaspectratio, scale=0.65]{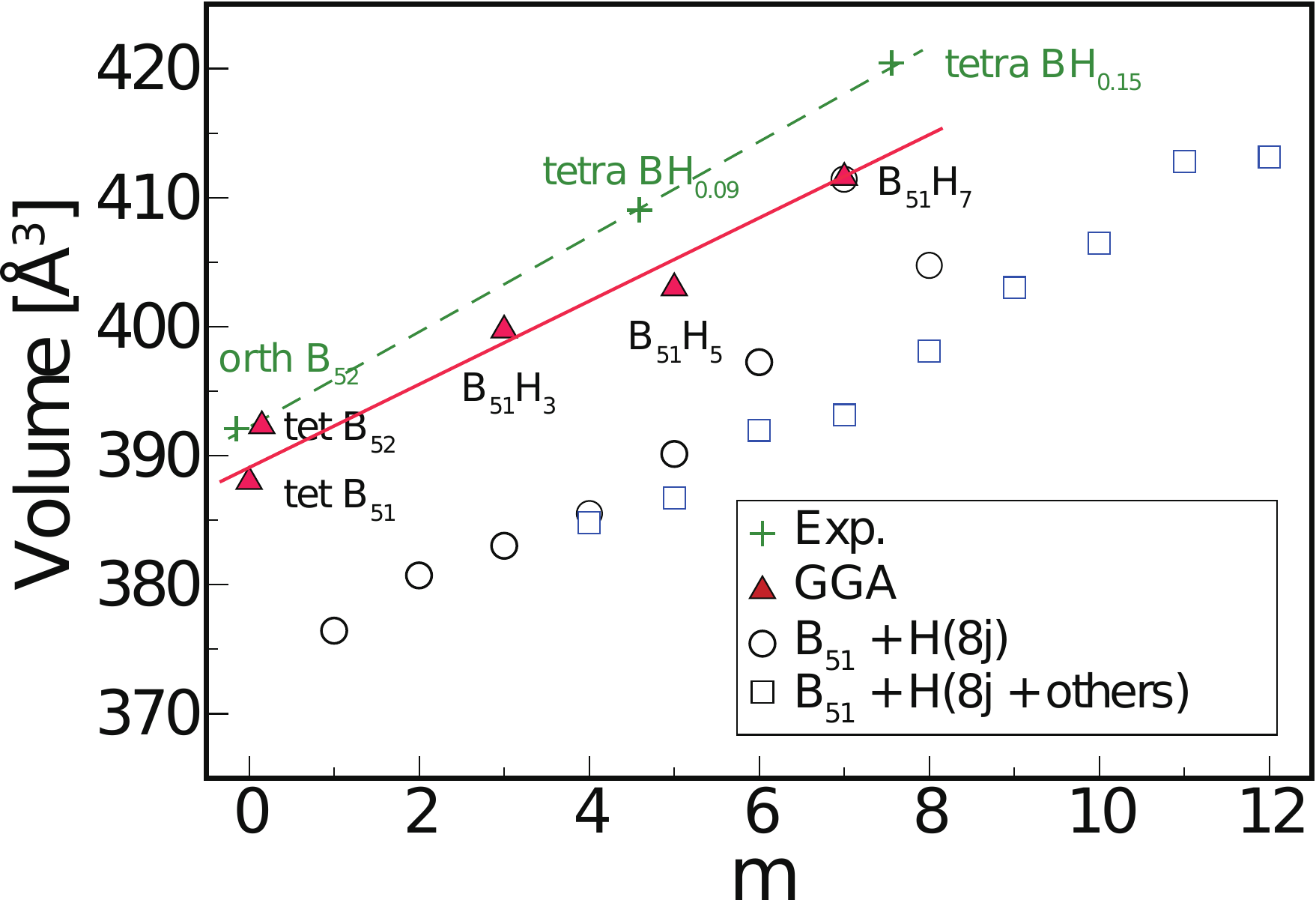}
      \caption{Increase of unit cell volume of B$_{n}$H$_{m}$ crystals with increasing hydrogen content. $m$ is the number of H atoms per cell. Open symbols correspond to DFT/LDA results, red triangles are DFT/GGA results, green crosses are experimental results from Ekimov et al.,\cite{Ekimov16} for the latter assuming 50, 51, and 52 boron atoms per cell.}
      \label{fig:volume}
\end{figure}

Figure \ref{fig:volume} shows the dependence of the cell volume on the H content $m$. All curves, experiment (green crosses)\cite{Ekimov16} and DFT calculations (open and filled symbols), nearly have the same slope and show a linear increase of the cell volume by about 10\% as the H content increases from $m=0$ to 12. We do not find an anisotropy in the increase between the $a$ and $b$ lattice directions. As usual, DFT/LDA calculations (open symbols) underestimate lattice parameters by about 1\% and the present calculations conform this trend. Using the GGA approximation instead (filled triangles) leads to much better
agreement with the experimental data. Thus Fig.~\ref{fig:volume} can be used to estimate the H content of hydrogenated boron.

\section{Two-step transition during dehydrogenation}
\label{sec:anneal}

In this section, we investigate the dehydrogenation of B$_{51}$H$_{m}$ crystals by annealing, as performed by Ekimov {\it et al.}\cite{Ekimov16} 
In their experiments the as-grown B$_{51}$H$_{7.7}$ crystals were subject to annealing at ambient pressure and a two-step phase transition was observed. In the first step the hydrogen content was reduced to $m=4.6$ at $T_{a1}=600-720$ K (the mean value is $660$ K). Further raising the temperature led to the complete release of H at $T_{a2}=820-970$ K (the mean value $900$ K).

\begin{figure}[htpb]
\centering
\includegraphics[keepaspectratio, scale=0.90]{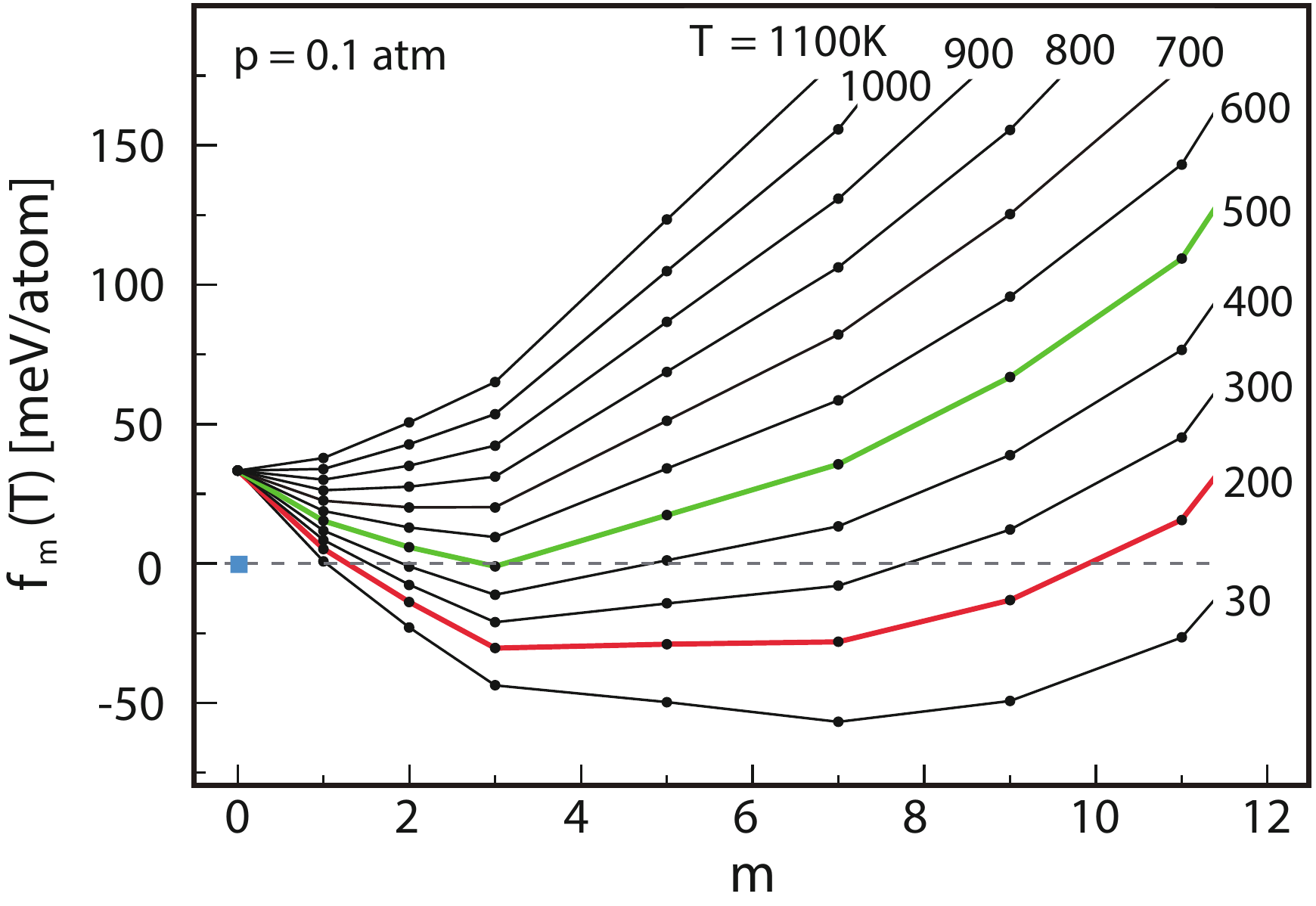}
\caption{Modeling the two-step transition during dehydrogenation. The figure shows the dependence of the free energy $f_{m}(T)$ of B$_{51}$H$_{m}$ crystals on the hydrogen-content $m$ and on temperature $T$. The energy zero is the lowest energy structure of pure $\alpha$-T B$_{52}$ (blue square). The partial pressure of hydrogen gas is $p=0.1$ atm. $f_{m}(T)$ includes the configurational entropy of crystal B$_{51}$H$_{m}$ as well the the $T$ dependence of the chemical potential of hydrogen. With increasing $T$, the minimum of $f_{m}$ (the thermodynamically most favorable structure) first changes from $m=7$ (B$_{51}$H$_{7}$) to $m=3$ (B$_{51}$H$_{3}$) at $T \approx 200$ K (red) and at $T \approx 500$ K (green), it changes to $\alpha$-T B$_{52}$. 
}
\label{fig:Fvsm}
\end{figure}

The partial free energy $f_{m}$ calculated by using Eqs.~(\ref{eq:muofH}) and (\ref{eq:formF}) is plotted in Fig.~\ref{fig:Fvsm}. In the experiment by Ekimov {\it et al.}, the pressure was not specified, hence we treat it as a parameter and use $p=0.1$ atm here. The data for even $m$ are omitted from this plot, because we already know that the compositions of even $m$ are not favorable.
%
In the low-temperature limit $f_{m}(T=30\ {\rm K})$ is essentially the same as the envelope of the lowest $e_f$ in Fig.~\ref{fig:B51H}. The free energy minimum is found for $m=7$ (B$_{51}$H$_{7}$) in good agreement with the experimental value of $m=7.7$, as already discussed above. 
Between $m=3$ to $7$, the free-energy curve $f_{m}$ is almost linear and as $T$ increases, the gradient $(\partial f_{m}/\partial m)_{T}$ gets more and more positive. This is because the chemical potential of hydrogen $\mu_{\rm H_{2}}(T,p)$, Eq.~(\ref{eq:muofH}), increases as $T$ increases. At about $T_{1}=200$ K, the free-energy curve $f_{m}$ is almost flat and the minimum changes abruptly to $m=3$.
As $T$ increases further, B$_{51}$H$_{3}$ remains the thermodynamically most favorable phase up to $T_{2} \approx 500$ K, where the free-energy minimum is found for B$_{52}$. 
These two transitions at $T_{1} \approx 200$ and $T_{2} \approx 500$ K seem to correspond to the two-step phase transition observed in the experiment, at $T_{a1}=660$ and $T_{a2}=900$ K, although the actual values are quite different.
We also examined the phonon contribution to the free energy $f_{m}$ (see Supplemental Material S.3.).
In the range $T<1100$ K it increases $f_m$(T) by $15 - 30$ meV/atom. But the difference between $f_{3}(T)$ and $f_7(T)$ due to phonons amounts to less than 5 meV/atom. This modifies $T_{1}$ and $T_{2}$ by less than 50 K.  We thus conclude that the phonon contribution will not lead to qualitative changes of this transition.

Although the theoretical transition temperatures are very different from the experimental values, our calculations clearly reproduce the two-step transition.
If H atoms were randomly incorporated in the crystal, there would be no sharp transition. The rate of hydrogen release would be a continuous function of temperature, namely, $\sim \exp(-Q/k_{\rm B}T)$, where $Q$ is the activation energy for the hydrogen release. 
The observation of relatively sharp transition temperatures indicates that the dehydrogenation process is a cooperative process that destroys long-range order. 
Here it is mostly related to the removal of H atoms from $(8j)$ and $(4g)$ sites.
Thus, our B-H system is essentially a compound with two stoichiometric compositions B$_{51}$H$_{7}$ and B$_{51}$H$_{3}$. However, in the experiment fractional compositions of B$_{51.5}$H$_{7.7}$ and B$_{51.5}$H$_{4.7}$ were found. For finite-temperature synthesis, there is additional degree of freedom, namely, the occupation of POS, which creates non-vanishing entropy, similar to ice crystals. Accordingly, deviations from the stoichiometric compositions can be expected at finite temperatures, which are likely to hide those clear stoichiometric compositions.

The removal of interstitial atoms from a compound by annealing is observed when there is an open channel for the interstitial atoms to move. This removal may be called ``thermal degassing".\cite{Kim15} For silicon clathrates, thermal degassing of sodium atoms was achieved by heating in vacuum. This process occurred at a certain temperature: $T=635$ K for Na$_{24}$Si$_{136}$ clathrate \cite{Gryko00,Stefanoski12} and $400$ K for Na$_{4}$Si$_{24}$ clathrates \cite{Kim15}. In these clathrates, sodium atoms occupy definite sites (implying long-range order) and therefore a sharp phase transition is observed. Sung {\it et al.} proposed the idea that new silicon allotropes might be obtained by degassing the foreign species from a clathrate compound.\cite{Sung18} 
The present study  extents the idea of thermal degassing to a more universal tool for material syntheses.

\section{Phase transition to orthorhombic structure}
\label{sec:phase-transition}
\subsection{Significance of orthorhombic distortion}

A key observation of the experiment by Ekimov {\it et al.} \cite{Ekimov16} is that, upon complete removal of hydrogen from hydrogenated $\alpha$-T boron, the tetragonal lattice becomes orthorhombic. We analyzed this $\delta$-O structure and found that it is only a slightly distorted variant of the parent hydrogenated $\alpha$-T boron crystals. See Supplemental Material S.4.
Prior to the work of Ekimov {\it et al.}~only $\alpha$-T crystals were experimentally reported. 
However, similar orthorhombic structures were theoretically studied by Hayami and Otani \cite{Hayami10} and later by Zhu {\it et al.}\cite{Zhu12} A detailed comparison of these three systems (see Supplemental Material S.4) revealed that they are identical, due to nearly vanishing structural and energetic differences.
Because of the excellent consistency between the experimental and the theoretical structures, we conclude that the $\delta$-O structure indeed exits. 
Then an important question is: why and how can the $\alpha$-T structure be transformed to the $\delta$-O structure?

\begin{table}
\begin{ruledtabular}
\caption{The strong correlation between the atomic configuration of partially occupied $(4c)$ boron sites and the orthorhombic lattice distortion $\Delta a/a$ within the primitive unit cell. 
Structures with tetragonal (T) and orthorhombic (O) symmetry are compared and their energy differences (O -- T) are given. 
The energy zeros of the formation energies $e_{f,m}$ for every group of systems are indicated in italics. 
For B$_{51}$H$_{m}$, the lowest-energy configuration are given. The atomic configuration of B$_{51}$H$_{3}$ is approximately out-of-plane.
The GGA functional was used.}
 \label{table:LD_primitive}
  \begin{tabular}{l l| D{.}{.}{-1} D{.}{.}{-1} D{.}{.}{-1} | D{.}{.}{3}}

   \multicolumn{2}{c|}{Structure} & \multicolumn{3}{c|}{$e_{f,m}$ (meV/atom)} &  \multicolumn {1}{c}{$\Delta a/a$ } \\ \cline{3-5} 
                                  &              & {\rm T} & {\rm O} &  {\rm O-T} &  \multicolumn {1}{c}{(\%)} \\  
\hline
    B$_{51}$    &  $1 \times (4c)$   & 31.0  & 30.8 & -0.2 & 0.21   \\ \hline
    B$_{52}$ & $2 \times (4c)$, in-plane    & 3.9    & 3.7   & -0.2 & 0.00 \\
                    & $2 \times (4c)$, out-of-plane      & \textit{0.0} & -1.7  & -1.7 & 1.81   \\ 
    \multicolumn{1}{r}{Zhu \cite{Zhu12}}  & $(4c) + (8h)$    & -1.0    & -1.5   & -0.5  & 1.08 \\ \hline
    B$_{51}$H$_{3}$ & $3 \times (8j)$  & -38.9  & -40.2 & -1.3& 1.44   \\ \hline
    B$_{51}$H$_{7}$ & $5 \times (8j)+ 2 \times (4g)$ & -41.8  & -43.2  & -1.4 & 0.29   \\ \hline
    \multicolumn {2}{l|}{Exp. B$_{51.5-52}$}                                      &                                    &                                                      & & 1.24   \\ \hline \hline

   B$_{50}$C$_{2}$ & in-plane  &  0.0 &  0.0 & 0.0 & 0.00     \\ 
                                & out-of-plane  & \textit{0.0} & -0.1 & -0.1 & 1.38     \\ 
 \hline
   B$_{50}$N$_{2}$  & in-plane  &  -0.1 &  -0.2 & -0.1  & 0.00     \\
                                 & out-of-plane  &  \textit{0.0} &  -0.1 & -0.1  & 1.08     \\

 \end{tabular}
\end{ruledtabular}
\end{table}

To address this question, let us explain why the lattice distortion is sensitive to the atom configuration of $(4c)$ boron sites.
As seen in Fig.~\ref{fig:zig-zag}, the out-of-plane configuration creates linear chains connecting  $(2b)$ and $(4c)$ sites in either the $a$- or $b$-direction. This implies the formation of strong covalent bonds in the direction of the chain that lead to the anisotropy of the orthorhombic lattice. The in-plane configuration, on the other hand, creates these linear chains in both directions and therefore no anisotropy exists. 
These simple geometrical insights are well reflected in the results shown in table \ref{table:LD_primitive}. 
It shows the calculated and experimental values for the magnitude of the lattice distortion $\Delta a/a$ and the calculated energy difference (O -- T) when the crystal undergoes a transformation T $\rightarrow$ O. 
%

The calculations show that a lattice distortion $\Delta a/a$ of more than 1\% only occurs for the out-of-plane configuration in the considered systems. The latter cannot be realized in B$_{51}$ because only one $(4c)$ site is occupied and this explains why the distortion is small. Further study by using supercell does not change this correlation (see Supplemental Material S.5).
This correlation between the lattice distortion and the atom configuration of POS's is also seen in the compounds. For B$_{51}$H$_{3}$, the calculated distortion is large, $\Delta a/a =1.44$ \%. This is reasonable, because the arrangement of three H atoms of B$_{51}$H$_{3}$ is approximately in the out-of-plane configuration. In the experiment, B$_{51}$H$_{3}$ has a T lattice, which suggests that H atoms are randomly distributed.
For B$_{52}$C$_{2}$ and B$_{52}$N$_{2}$ similar large distortions of more than 1 \% are calculated for the out-of-plane configuration. However, experimentally these crystals are found to be tetragonal. In these compounds, the two interstitial B atoms are located at $(8h)$ or $(8i)$ sites, and are likely to be occupied randomly.\cite{Amberger71, Ploog72, Will76} 
Therefore, the presence of the O lattice in $\delta$-O boron is compelling evidence for the ordering of the two $(4c)$-site atoms in the out-of-plane configuration.

In our previous paper, we pointed out that the out-of-plane configuration leads to geometrical frustration because a perfect anti-ferromagnetic (AF) arrangement of FOS and POS is symmetry incompatible with the perfect order in the out-of-plane configuration.\cite{Uemura16} However, this symmetry incompatibility can be relaxed, if the lattice is distorted to the O lattice. Although this distortion does not eliminate in-gap states 
it slightly reduces the instability caused by the frustration.

To summarize this subsection, $\delta$-O boron is the ordered variant of $\alpha$-T boron with respect to the occupation of the $(4c)$ sites with boron.
The orthorhombic distortion occurs when two conditions are met: (i) the stoichiometric composition B$_{52}$ is present, (ii) the $(4c)$ sites are occupied according to the out-of-plane configuration.

\subsection{Transition mechanism: an order-disorder transition by cooperative atom migration}

By bearing in mind the significance of the orthorhombic distortion, let us now discuss the behavior of the interstitial B atoms at the T$ \rightarrow $O transition.
One unusual fact of the experiment is that the B content $n$ apparently increases from 51.5 to 52 during the transition.  This increase is consistent with our finding that the orthorhombic distortion can only occur for $n=52$. 
However, where the extra B atoms come from is unclear.
One possibility may be the agglomeration of B atoms during the initial crystal growth, that could provide mobile B atoms.

Another puzzling point is the apparent ordering of POS during the annealing process.
The fact that the parent crystal B$_{51}$H$_{7}$ has tetragonal symmetry is consistent for multiple reasons. First, since the energy difference of different H configurations is small, a random occupation is to be expected for high-$T$ synthesis. 
Second, the non-stoichiometrc composition of $n=51.5$ also supports the tetragonal symmetry.\cite{Uemura16} 
Third, the gain in the formation the energy for the T $\rightarrow$ O transformation (as shown in Table \ref{table:LD_primitive}) is also very small (less than 2 meV/atom). This is on the edge of the accuracy of the GGA functional. Thus the T and O structures may be considered as quasi-degenerate. In this case the the free energy at high temperatures is dominated by entropic contribution and therefore a random occupation of the $(4c)$ interstitial B atoms is very likely to occur.
On the other hand, it is certain from the discussion above that the $(4c)$ interstitial B atoms are ordered in the out-of-plane configuration after the T$ \rightarrow $O transition.
Because the structural difference between $\delta$-O and $\alpha$-T boron is only the order of the two $(4c)$-site atoms, this transformation is an order-disorder transition.  
Such a transition is suggested theoretically by Widom and Huhn for boron-carbide, although so far there is no experimental evidence.\cite{Widom12,Huhn13} 
For metallic alloys, order-disorder transitions are well-known: for example, CuZn alloys exhibit a transition at $T_{\rm od}=465^{\circ}$C, which are associated with the structural transformation from BCC to a CsCl-type structure.
\cite{Shewmon} But, it is rare to observe an order-disorder transition in strongly bound, covalent crystals.

\begin{figure}[tp]
\includegraphics[width=10.2 cm]{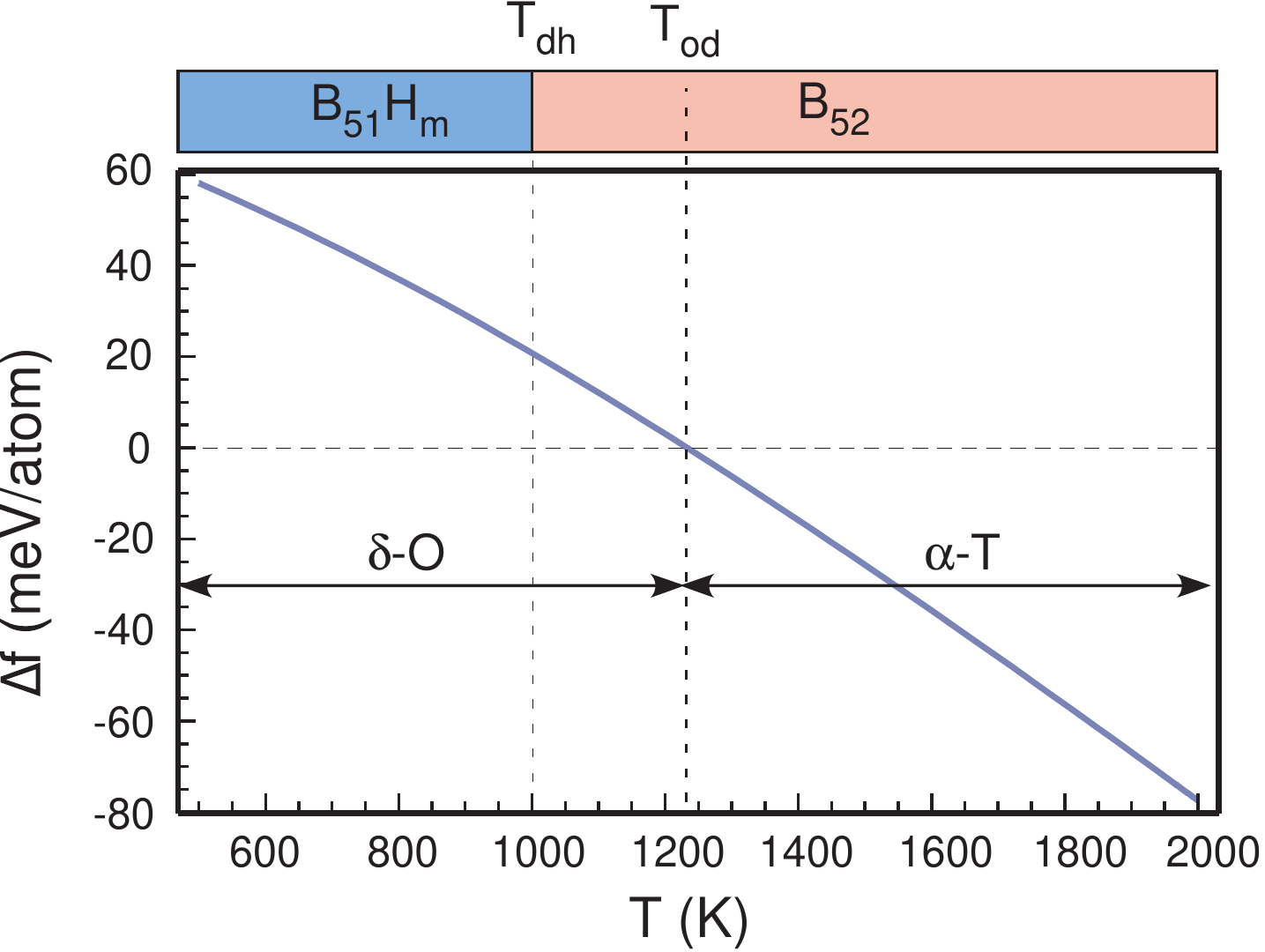}
\caption{The appearance of an order-disorder transition at $T_{\rm od}$ as shown by the difference of the free energy $\Delta f$ between $\delta$-O and $\alpha$-T B$_{52}$. Thermodynamically favorable phases are shown by double arrows, the temperature of complete dehydrogenation $T_{\rm dh}$ is also indicated. It is proposed that the dehydrogenation facilitates the structural transition by cooperative migration of B and H atoms.
} 
\label{fig:O-Dtransition}
\end{figure}

%

We modeled the order-disorder transition with a free-energy model. The details of the calculation are given in Appendix A. As shown in Fig.~\ref{fig:O-Dtransition}, the order-disorder transition temperature $T_{\rm od}$ is 1260 K.
This phase transition identifies the ordered $\delta$-O boron to be the low-temperature modification and $\alpha$-T boron to be the high-temperature modification of B$_{52}$; the latter is stabilized by the configurational entropy of disordered POS. 
Therefore, it is no surprise that only $\alpha$-T crystals were obtained in all the HPHT syntheses reported, so far. For example, Qin {\it et al.}~synthesized $\alpha$-T boron at  $T > 1800 ^{\circ}$C \ \cite{Qin12}. 
Although the temperature range of $T=1100 - 1300^{\circ}$C is lower than those of Qin {\it et al.}, the experiment by Ekimov {\it et al.} is still in the high temperature range, so that obtaining of $\alpha$-T phase is reasonable. This high-temperature phase, however, persists even at low temperatures as a {\it frozen-in} state, because it is difficult for atoms to move once a material cooled to ambient conditions.

Figure \ref{fig:O-Dtransition} finally reveals the puzzles related to the transition. One of them is that two different types of transitions, hydrogen release and the ordering of interstitial atoms, apparently occur simultaneously.  At first glance, the two seem to be unrelated and therefore should occur at different temperatures $T_{\rm dh}$ (or $T_{a2}$ in the annealing experiment) for complete dehydrogenation and $T_{\rm od}$ for the order-disorder transition. And indeed Fig.~\ref{fig:O-Dtransition} shows that $T_{\rm dh}$ is lower than $T_{\rm od}$. Theoretically, a crystal should transform between O and T at $T_{\rm dh}$. 
But, migration of atoms in solids does not occur at low temperatures, due to their low mobility. Therefore this order-disorder transition was not observed before.
Here we propose that the role of dehydrogenation is to stimulate the migration of B atoms through the motion of H atoms upon release. Then the temperature difference $T_{\rm od}-T_{\rm dh}$ is the driving force of the transition.
Moving B atoms may not be at all ruled out for boron-rich crystals; in $\beta$-rhombohedral boron, it is currently speculated that B atoms migrate at a moderate temperatures.\cite{Hoffmann12, Ogitsu13}. Hence, it is possible that the motion of H atoms enhances the migration of B atoms.

Another puzzle is the process of the ordering.
To obtain the ordered out-of-plane configuration from B$_{51}$H$_{3}$ during annealing, not only new B atoms but also atoms at $(4c)$ sites must migrate. This could be difficult because these atoms already form strong covalent bonds with their neighbors. However, $(4c)$ sites are interstitial sites. 
In silicon the diffusion of impurities is enhanced by accompanying migration of intrinsic defects, vacancies or self interstitials.\cite{Fahey89} The H sites in B$_{51}$H$_{3}$ are close to $(4c)$ B sites, so that removing a H atom at a $(8j)$ site is likely to affect the occupation of a nearby $(4c)$ B atom.
Therefore, we hypothesize that the T$\rightarrow$O transition occurs because the migration of B atoms is assisted by the migration H atoms upon hydrogen release.
So the three types of changes, the evacuation of H atoms, the migration of interstitial B atoms, and the lattice distortion T$\rightarrow$O work cooperatively.



Lastly, we surmise about another cooperative role that hydrogen might have during the growth process.
As stated above, the temperature of $1100 - 1300^{\circ}$C, at which Ekimov {\it et al}.~synthesized hydrogenated $\alpha$-T boron, is relatively low compared with the one of  Qin {\it et al.}
It may be possible that hydrogen assists the nucleation of $\alpha$-T boron at relatively low temperatures. The role of hydrogen on crystal growth was studied by many authors for the epitaxial growth of semiconductors. Here we only refer to a few works.\cite{Sugaya91,Okada95,Kawabe95,Notzel99,Shimizu05}

\section{Conclusion}
\label{sec:conclusion}
We have studied the structure and thermodynamic stability of hydrogenated $\alpha$-tetragonal boron, its dehydrogenation process and the transition to $\delta$-orthorhombic boron B$_{52}$ with density functional theory calculations.
Our results show that hydrogenated $\alpha$-tetragonal boron has at least two stable stoichiometric compositions, B$_{51}$H$_{7}$ and B$_{51}$H$_{3}$, where the hydrogen atoms are mostly occupying $(8j)$ sites and, to a minor extend, $(4g)$ sites. 
Owing to the presence of partially occupied sites, deviations from these ideal stoichiometric composition are to be expected for crystals synthesized at high pressure and high temperature.
By thermodynamic modeling we could qualitatively reproduce the two-step phase transition reported by Ekimov {\it et al.}~upon annealing.\cite{Ekimov16} It  corresponds to successive transitions from B$_{51}$H$_{7}$ to B$_{51}$H$_{3}$ to pure B$_{52}$. 
The so obtained $\delta$-orthorhombic boron is an ordered, low-temperature phase and $\alpha$-tetragonal boron is a disordered, high-temperature phase of B$_{52}$. The two are connected by an order-disorder transition, that is associated with the ordering of interstitial boron atoms at (4c)-sites. 
This ordering implies boron atoms to migrate, which is usually hindered in a strongly bound, covalent crystal. Our analysis reveals that the migration of boron atoms is likely to be assisted by the migration of hydrogen atoms upon annealing and we refer to this mechanism as "cooperative atom migration".

Our results represent an important step forward for the understanding of boron and hydrogenated boron crystals. We determined the atomic structure of stable hydrogenated boron compounds and revealed structural and thermodynamic relations between  $\alpha$-tetragonal boron and $\delta$-orthorhombic boron. These findings are also in excellent agreement with the experiments of Ekimov {\it et al.} \cite{Ekimov16}
The present study opens an new avenue to control or remove the intrinsic defects of covalently bound crystals by utilizing volatile, foreign atoms. This was previously considered to be difficult.
\begin{acknowledgments}
N.U.~thanks Dr.~T.~Ogitsu (Lawrence Livermore National Lab.)~for his hospitality to the visit of N.U.~to his laboratory and for valuable discussions. This work is financially supported by the Innovative Professional Development project of Professional development Consortium for Computational Materials Scientists. K.S.~received a financial support by the Research Program of ``Five-star Alliance" in ``NJRC Mater.~\& Dev."
J.K. acknowledges funding by the German Research Foundation (DFG) under grant number SE 651/45-1 and financial support by the Center for Advancing Electronics Dresden.
\end{acknowledgments}

\appendix

\section{Order-disorder transition temperature}
For random alloys, order-disorder transitions are treated by the Bragg-William model.\cite{Fowler-Guggenheim} Here the transition is related to the ordering of (4c)-site B atoms, where $\delta$-O boron is the ordered state and $\alpha$-T is the disordered one. Therefore, the transition temperature $T_{\rm od}$ can be obtained directly by calculating their free energy difference
$\Delta F_{\alpha,\delta} = F_{\alpha-{\rm T}} - F_{\delta-{\rm O}}$. It is given by
\begin{equation}
\Delta F_{\alpha,\delta} = F_{\alpha} = E_{\alpha,0} -\frac{1}{\beta} \ln \left[ \sum_{i} g_{\alpha,i} e^{-\beta \Delta E_{\alpha,i}} \right] ,
\label{eq:append-2}
\end{equation}
where $g_{\alpha,i}$ is the multiplicity of $i$-th energy level of $\alpha$-T boron, assuming that the entropy difference between $\alpha$-T boron and $\delta$-O boron comes entirely from the configuration entropy of $\alpha$-T boron $S_{\alpha}^{\rm c}$. 
In Eq.~(\ref{eq:append-2}), $E_{\alpha,0}$ is the ground-state energy of $\alpha$-T boron relative to that of $\delta$-O boron, which is 1.7 meV/atom from Table \ref{table:LD_primitive}. $\Delta E_{\alpha,i}$ is the $i$-th energy level of $\alpha$-T boron relative to $E_{\alpha,0}$. Here, the summation $i$ is taken over $i=0, 1$, which correspond to the out-of-plane and the in-plane configurations, respectively.
The temperature satisfying $ \Delta F_{\alpha,\delta} =0$ gives the order-disorder transition temperature $T_{\rm od}$; we obtained $T_{\rm od}=1260$ K.
The free energy difference  is shown in Fig.~\ref{fig:O-Dtransition} as a function of $T$. 

\bibliography{boron,added}


\end{document}